\documentclass[%
 reprint,
 amsmath,amssymb,
 aip,
 jcp
]{revtex4-2}

\usepackage{graphicx}
\usepackage{dcolumn}
\usepackage{bm}
\usepackage[dvipsnames]{xcolor}

\begin{document}

\title{Extending and validating bubble nucleation rate predictions in a Lennard-Jones fluid with enhanced sampling methods and transition state theory}

\author{Kristof M. Bal}
  \email{kristof.bal@uantwerpen.be}
  \affiliation{Department of Chemistry and NANOlab Center of Excellence, University of Antwerp, Universiteitsplein 1, 2610 Antwerp, Belgium}
\author{Erik C. Neyts}
  \affiliation{Department of Chemistry and NANOlab Center of Excellence, University of Antwerp, Universiteitsplein 1, 2610 Antwerp, Belgium}

\date{26 September 2022}

\begin{abstract}
We calculate bubble nucleation rates in a Lennard-Jones fluid through explicit molecular dynamics simulations.
Our approach---based on a recent free energy method (dubbed reweighted Jarzynski sampling), transition state theory, and a simple recrossing correction---allows us to probe a fairly wide range of rates in several superheated and cavitation regimes in a consistent manner.
Rate predictions from this approach bridge disparate independent literature studies on the same model system.
As such, we find that rate predictions based on classical nucleation theory, direct brute force molecular dynamics simulations, and seeding are consistent with our approach and one another.
Published rates derived from forward flux sampling simulations are, however, found to be outliers.
This study serves two purposes.
First, we validate the reliability of common modeling techniques and extrapolation approaches on a paradigmatic problem in materials science and chemical physics.
Second, we further test our highly generic recipe for rate calculations, and establish its applicability to nucleation processes.
\end{abstract}


\maketitle

\section{Introduction}
Nucleation processes are important throughout nature and technology, and are therefore a long-standing research area within the physical and chemical sciences. 
Even so, many uncertainties remain around our theoretical understanding of nucleation mechanisms.~\cite{Karthika2016}
Atomistic simulations are, in principle, ideally suited to study nucleation with a high level of detail.
Accurate, physically meaningful simulation setups are however challenging to construct with many, sometimes subtle, possible sources of error.~\cite{Blow2021}

The difficulties associated with nucleation simulations are nicely illustrated by a seemingly simple and rather well-defined system: Homogeneous bubble nucleation in a Lennard-Jones (LJ) fluid.
Different modeling studies disagree about the nucleation mechanism,~\cite{Wang2009,Diemand2014} the nucleation rate,~\cite{Wang2009,Meadley2012,RosalesPelaez2019} the validity of classical nucleation theory (CNT) for the process,~\cite{Wang2009,Meadley2012} and the assumptions underpinning existing CNT models.~\cite{Tanaka2015,Schmelzer2016}
These literature studies employed different methodologies, such as direct brute force molecular dynamics (MD),~\cite{Diemand2014,RosalesPelaez2019} forward flux sampling (FFS),~\cite{Wang2009,Meadley2012} or seeding~\cite{RosalesPelaez2019} which have been compared to CNT predictions~\cite{Diemand2014,Tanaka2015,Schmelzer2016}
Due to different time and length scale restrictions of these methods, different physical conditions were probed, resulting in limited overlap between rate data.
A rigorous cross-validation of these methodologies, and their respective rate predictions, is therefore mostly lacking.

Recently, we proposed a generic strategy based on transition state theory (TST) to evaluate rates of diverse processes.~\cite{Bal2020,Bal2021JCP}
A key advantage of this strategy is that it can fully leverage the rich toolbox of enhanced sampling methods for free energy calculation.~\cite{Henin2022}
As a result, very wide time scale ranges can be studied within a consistent simulation strategy.~\cite{Bal2021JCP}

In this work, we revisit the bubble nucleation in LJ fluids using this methodology, simultaneously verifying existing rate predictions, and closing the gap between disparate literature conditions.

\section{Methodology}
\label{sec:method}

Rate estimates for bubble nucleation are obtained in a system- and process-agnostic manner based on recent developments in the field of enhanced sampling approaches.~\cite{Bal2020,Bal2021JCP,Bal2021JCTC}
The approach employs molecular dynamics simulations in relatively small simulation cells to ultimately yield macroscopic nucleation rates over a wide range of conditions and time scales.
The methodology summarized here has already been successfully applied to droplet nucleation from supersaturated vapor in a Lennard-Jones system.~\cite{Bal2021JCP,Bal2021JCTC}

\subsection{Free energy calculation in a finite simulation cell}

Nucleation is a rare event and therefore, in general, difficult to observe in molecular simulations.
One way to overcome the time scale problem, is the application of a bias potential $V$.
Such a bias, if properly designed, allows to sample the sample critical bubbles as well as metastable liquid states within the same simulation, so that a nucleation free energy surface (FES) can be constructed from the (reweighted) marginal probability density along the nucleation path.

In the specific approach used here, dubbed reweighted Jarzynski sampling,~\cite{Bal2021JCTC} the bias potential is generated from a small number of nonequilibrium simulations in which the system is pushed from the metastable liquid towards states beyond the the critical bubble; a putative nucleation free energy surface $\widetilde{G}$ is learned from the associated nonequilibrium work distribution by an approximation of the Jarzynski equality.~\cite{Jarzynski1997}
The final free energy estimate $G$ is then obtained from longer sampling runs under influence of the bias potential $V = -\widetilde{G}$.

For any collective variable $\chi (\mathbf{R})$ that is a function of the system coordinates $\mathbf{R}$, the free energy $G(\chi)$ at a constant temperature $T$ and pressure $p$ is defined as
\begin{equation}
  G(\chi) = - k_B T \ln P(\chi) , \label{eq:FES}
\end{equation}
in which $k_B$ is the Boltzmann constant and $P(\chi)$ the marginal probability distribution of $\chi$ under the considered conditions.
$P(\chi)$ can be sampled in a biased simulation by accumulating the histogram for $\chi$ while reweighting each sampled data point $\chi_i$.
The weight $w_i$ of each sampled data point $i$ in the histogram is given by the umbrella sampling relation $w_i = e^{V_i/k_B T}$, with $V_i$ the instantaneous value of a time-independent $V$.~\cite{Torrie1977}

Reweighted Jarzynski sampling addresses the exploration--convergence conundrum inherent to adaptive bias enhanced sampling methods.~\cite{Invernizzi2022}
When using imperfect collective variables, a single sampling run can either be optimized to target many transitions (exploration) or a stable free energy estimate (convergence).
This issue is here essentially sidestepped by splitting up the simulation in a forced exploration phase with a nonequilibrium bias and a convergence phase with a non-adaptive bias.

An advantage of the outlined reweighting strategy is that $V$ need not be a function of $\chi$.
That is, biasing and sampling can be performed on different collective variables.
This is precisely what we will take advantage of in this work.
Most nucleation studies rely on reaction coordinates that implement some variant of the ten Wolde--Frenkel parameter $n$,~\cite{tenWolde1998} which counts the number of atoms that are part of the nucleating phase.
For bubble nucleation, this means that $n$ should count the number of vapor atoms.
We use a continuous expression for $n$ based on geometric switching functions to count the number of atoms with less than five neighbors closer than $1.6\sigma$.~\cite{Wang2009,Meadley2012}
This manybody expression for $n$ is however very expensive to evaluate, especially if also atomic gradients are needed to apply a bias potential $V(n)$.
Therefore, we perform initial nonequilibrium simulations by biasing the average molar volume $v_m$ in the system, yielding $\widetilde{G}(v_m) = - V(v_m)$.
The final nucleation FES $G(n)$ is then calculated from a subset of the configurations sampled in the biased simulation under influence of $V(v_m)$.

It must be noted that our implementation of $n$ is evaluated for all atoms in the system, and thus counts all vapor-like atoms.
This means that it is not strictly the parameter defined by ten Wolde and Frenkel---according to CNT only the largest cluster will drive the nucleation process, and $n$ should only count the atoms in this nucleus.
As previously discussed, we only require $n$ to be a sufficiently good reaction coordinate that allows us to distinguish pre- and postcritical states and parametrize a dividing surface $n=n^*$ (as described in the next section).~\cite{Bal2021JCP}
Such a set-up is consistent with previous enhanced sampling studies of droplet nucleation, where droplet growth was analyzed and driven by a parameter $n$ that counts all liquid-like atoms, defined as atoms with more than five neighbors.~\cite{Salvalaglio2016,Tsai2019,Bal2021JCP}

\subsection{Rates from transition state theory}

Transition state theory offers a theoretical framework for the calculation of rare event rates.
If we wish to employ the Eyring formulation of the TST rate,
\begin{equation}
  k^\mathrm{TST} = \frac{k_B T}{h} e^{-\Delta^\ddagger G / k_B T} , \label{eq:Eyring}
\end{equation}
the free energy barrier $\Delta^\ddagger G$ must be defined as~\cite{VandenEijnden2005,Bal2020}
\begin{equation}
  \Delta^\ddagger G = G(n^*) + k_B T \ln \frac{\langle |\nabla n| \rangle^{-1}_{n=n^*}}{h} \sqrt{2 \pi m k_B T} - G_l, \label{eq:barG} 
\end{equation}
with
\begin{equation}
  G_l = -k_B T \ln \int_{n < n^*}  \mathrm{d} n \, e^{-G(n) / k_B T} . \label{eq:Gl}
\end{equation}
In these equations, $h$ is the Planck constant, $m$ the mass of the particles, and $\langle |\nabla n| \rangle^{-1}_{n=n^*}$ the average norm of the gradient of $n$ with respect to all atomic coordinates at $n=n^*$.
The TST rate measures the total flux through the dividing surface $n=n^*$, which is always an upper bound to the effective rate of interest.
Therefore, the value of $n^*$, i.e., the location of the transition state, can be found by maximizing $\Delta^\ddagger G$ and, hence, minimizing the rate.

The key assumption underpinning TST is that the candidate reaction coordinate (here $n$) can correctly parametrize the dividing surface.
This idea is connected to the notion that $n$ is the slowest degree of freedom relevant to the transition.
A proper choice of the reaction coordinate is therefore critical.
It is possible to optimize the definition of the reaction coordinate itself by maximizing its time scale separation with other degrees of freedom.~\cite{Tiwary2016}
Such an optimized reaction coordinate, which also contains information on the nucleus shape, was developed in the context of droplet nucleation.~\cite{Tsai2019}
The droplet analogue of $n$, when plugged into our procedure, however still turned out to yield accurate rate estimations, which is why we also employ it here.~\cite{Bal2021JCP}

Two factors still separate the TST rate $k^\mathrm{TST}$ from the target macroscopic nucleation rate $J$:
\begin{enumerate}
  \item The FES $G(n)$ and rate $k$ are only defined for the specific simulation model, i.e., only within a small periodic simulation cell with a few thousand atoms;
  \item The TST rate contains all crossings of the dividing surface, even those that do not result into effective state-to-state transitions, which means that $k \leq k^\mathrm{TST}$.
\end{enumerate}

\subsection{Obtaining the final macroscopic rate}

The relation between the per-cell nucleation rate $k$ and macroscopic per-volume rate $J$ is given by
\begin{equation}
  J = \frac{k}{V} ,
\end{equation}
provided that the cell volume $V$ is large enough to avoid self-interaction of the nucleating bubble across periodic images.

Before arriving at $J$, we must first obtain $k$.
The relation between $k$ and $k^\mathrm{TST}$ can be expressed as
\begin{equation}
  k = \kappa k^\mathrm{TST} ,
\end{equation}
in which the transmission coefficient $\kappa$ accounts for recrossings of the TST dividing surface.

We have recently proposed a simple strategy for the determination of $\kappa$ that fits in the workflow of our free energy calculation.~\cite{Bal2021JCP}
In order to ascertain the ability of an approximate reaction coordinate to properly discriminate between the dividing surface, and the (meta)stable states on that it separates, one can use committor analysis of the putative dividing surface $n = n^*$.
As part of this committor analysis we prepare a number of equilibrated configurations confined at $n = n^*$ by using restraints and monitor their evolution after restraints are lifted.
The committor $p_g$ is the fraction of trajectories that results in a successful nucleation event, i.e., further growth of the gas phase.

Committor analysis thus serves as an \textit{a posteriori} validation of the employed order parameter.
A necessary condition for $n=n^*$ to be a dividing surface for bubble nucleation is observing $p_g \approx 0.5$.

If $n=n^*$ passes the committor test (so $p_g = 0.5$), additional insights can be extracted from the collection of committor trajectories.
The average number of crossings of the dividing surface $\langle j_\mathrm{cross} \rangle$ is correlated with $\kappa$.
If we assume that recrossings are the only contributor to $\kappa$, we have, by definition:
\begin{equation}
  \kappa = \frac{1}{2 \langle j_\mathrm{cross} \rangle} . \label{eq:committor}
\end{equation}
We have previously speculated that this definition of $\kappa$ might also compensate deficiencies in the chosen reaction coordinate $n$, provided $n$ does not deviate too much from the true reaction coordinate.~\cite{Bal2021JCP}

\section{Computational details}

All simulations were carried out with LAMMPS~\cite{Plimpton1995,Thompson2022} and the PLUMED plugin.~\cite{Tribello2014,PLUMED2019}
Machine learning algorithms were used as implemented in the scikit-learn library.~\cite{Pedregosa2011} 

Pairwise interatomic interactions were described using a truncated force-shifted Lennard-Jones (TFS--LJ) potential
\begin{equation}
  U(r) = \phi(r) - \phi(r_c) - (r-r_c) \left | \frac{d \phi (r)}{dr} \right |_{r=r_c} ,
\end{equation}
in which $\phi(r)$ is the standard Lennard-Jones potential of the interatomic distance $r$:
\begin{equation}
  \phi(r) = 4 \epsilon \left [ \left(\frac{\sigma}{r}\right)^{12} - \left(\frac{\sigma}{r}\right)^6 \right ] .
\end{equation}
We set $\sigma = \epsilon = k_B = m = 1$, thus using reduced units throughout.
The cutoff distance was $r_c = 2.5$.
These choices allowed us to compare computed rates to literature simulations based on an identical potential.
More specifically, we revisited literature work on the $T = 0.7$ and $T = 0.855$ isotherms,~\cite{Wang2009,Meadley2012,Diemand2014} and the $p = 0.026$ isobar of the same TFS--LJ fluid.~\cite{Meadley2012,Diemand2014,RosalesPelaez2019}
Sampling of the isobaric--isothermal (NpT) ensemble was performed with an isotropic Nosé--Hoover style thermo- and barostat,~\cite{Martyna1994} integrated with a 0.005 time step.
Initial thermalization used a Langevin thermostat.~\cite{Bussi2007}

Three distinct types of simulation were carried out for each condition.

In the first step, a set of 10 steered MD (SMD) simulations were carried out to obtain a nonequilibrium work distribution.
A moving harmonic restraint was used to bring $v_m$ from a low value (close to that of the liquid in equilibrium) to high value (beyond the critical nucleus) over $2 \times 10^6$ MD steps.
The range of $v_m$ was determined for each set of simulations on a trial-and-error basis.
The nonequilibrium work $W$ was recorded at regular time steps.
For each trajectory $i$, the noisy collection of $(v_m(t), W(t))$ data mapped to a smooth $W_i(v_m)$ curve using kernel ridge regression with a regularization strength of $10^{-3}$.
Using the cumulant expansion of the Jarzynski equality,~\cite{Park2004} the full set of $W_i$ curves was then finally converted into $\widetilde{G}(v_m)$, which was then fitted to an artificial neural network (ANN) consisting of a single 12-neuron hidden layer.

The second step involved five longer equilibrium sampling runs using the ANN bias $V(v_m) = -\widetilde{G}(v_m)$.
Sampling efficiencies were improved by limiting the simulation to a narrow range of $v_m$, based on the shape of $\widetilde{G}(v_m)$; harmonic restraints were used to prevent the system from moving too far beyond the critical nucleus.
The final free energy surface $G(n)$ was then obtained from the reweighted histogram $P (n)$ of $n$ according to Eq.~\eqref{eq:FES}.
Sufficient sampling was found to be possible within $2 \times 10^7$ MD steps per simulation.
The histogram was accumulated using the kernel density estimation functionality within PLUMED, and written to a grid.

The third  and final step of the procedure was committor analysis---after evaluation of the barrier Eq.~\eqref{eq:barG}, TST rate Eq.~\eqref{eq:Eyring}, and identification of the approximate dividing surface $n=n^*$.
SMD simulations were used to prepare 10 configurations at $n=n^*$ over $2 \times 10^6$ MD steps each.
The subsequent committor trajectories were analyzed to verify that $p_g \approx 0.5$, and $\langle j_\mathrm{cross} \rangle$ was then also evaluated.
In this step the $n$ CV was biased.
A multiple time stepping scheme~\cite{Ferrarotti2015} was used with a stride of 10 to limit the number of expensive force evaluations on $n$.

Representative input decks that contain the full set of simulation parameters and implementation details are available on PLUMED-NEST (www.plumed-nest.org), the public repository of the PLUMED consortium~\cite{PLUMED2019}, as plumID:22.025.\cite{data}

\section{Results and discussion}

\subsection{Free energy surface and choice of CV}

\begin{figure*}[tb]
\includegraphics{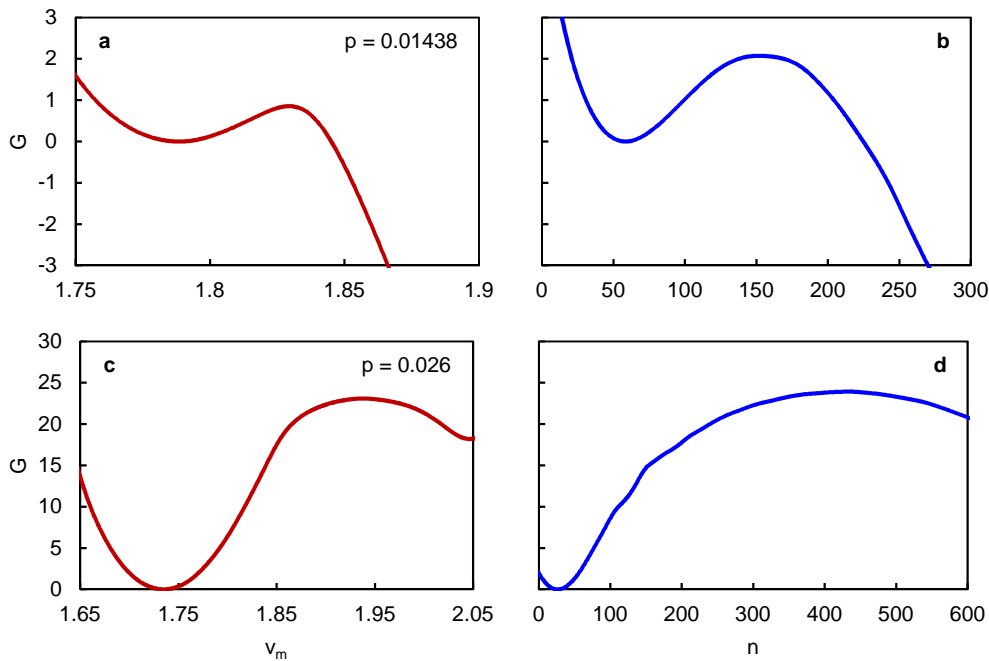}
\caption{\label{fig:panel-fes} Bubble nucleation FES at (a--b) $p = 0.01438$ and (c--d) $p = 0.026$ on the $T = 0.855$ isotherm of the TFS--LJ fluid, projected on $v_m$ and $n$, respectively.
Reduced LJ units are used throughout, meaning that $G$ is in units of $\epsilon$.}
\end{figure*}

We have used two CVs in our simulation: The inexpensive molar volume CV $v_m$ for biasing, and the expensive manybody CV $n$ for evaluation of the final barrier.
The effectiveness of this choice hinges on the assumption that a FES projected on $v_m$ sufficiently captures the barrier on $G(n)$.
That is, we assume that although $n$ is the best possible reaction coordinate and the true barrier can be evaluated from $G(n)$, the barrier on $G(v_m)$ only misses the true barrier by a few $k_B T$.
$V(v_m) = - G(v_m)$, while imperfect, therefore ought to be a perfectly cromulent bias to allow for frequent barrier crossing and good sampling.

We illustrate the applicability of our approach for two conditions on the $T = 0.855$ isotherm.
They represent a low barrier ($p = 0.01438$, Figure~\ref{fig:panel-fes}a--b) and a higher barrier case ($p = 0.026$, Figure~\ref{fig:panel-fes}c--d).
In both systems, the local maximum of $G(v_m)$ is lower than that of $G(n)$.
The relative valley-to-peak height of the FES maximum is not, strictly speaking, the free energy barrier, although it is usually quite close.~\cite{Bal2020}
In any case, it might be inferred that $n$ is a somewhat better reaction coordinate because it discriminates more sharply between metastable states and is the order parameter with the largest spectral gap.~\cite{Tiwary2016}
In absolute terms, however, the difference remains in the order of 1--2~$k_B T$.
A bias based on $G(v_m)$ can therefore still be successful.

\begin{figure}[tb]
\includegraphics{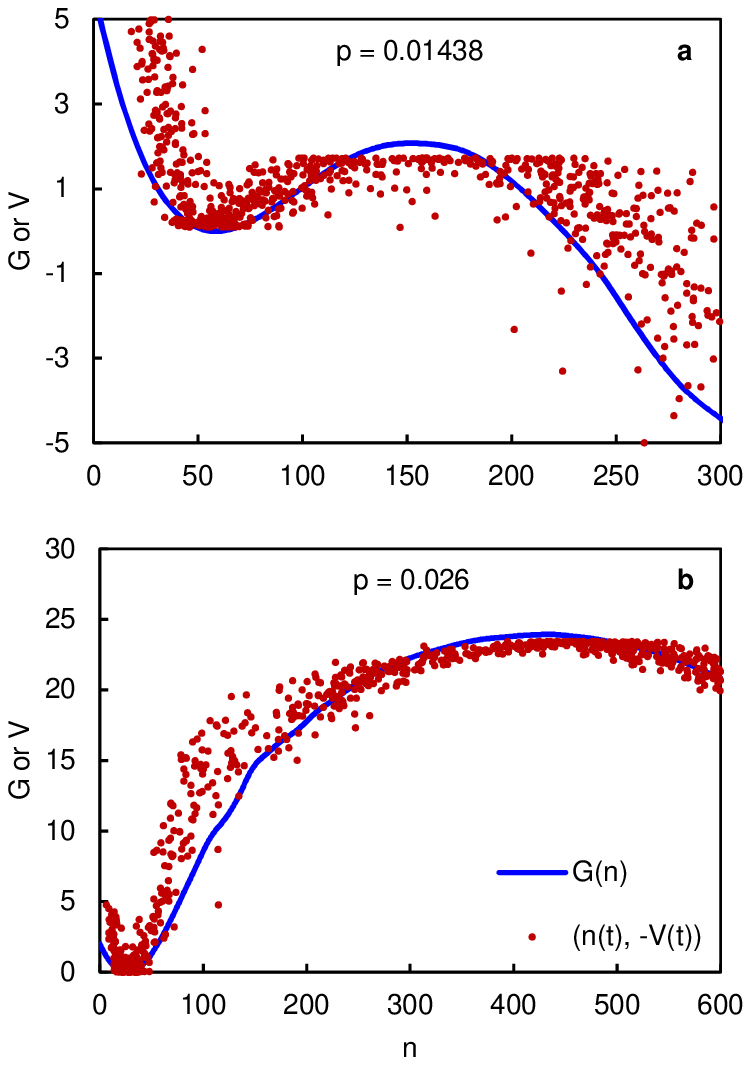}
\caption{\label{fig:panel-bias} FES $G(n)$ and values of the inverted bias $-V(v_m) = \widetilde{G}(v_m)$ recorded at different times $t$, plotted at $n(t)$ for (a) $p = 0.01438$ and (b) $p = 0.026$.}
\end{figure}

The bias $V(v_m)$ is based on the approximate $\widetilde{G}(v_m)$, rather than the true $G(v_m)$.
Both $n$ and $v_m$ are function of $\mathbf{R}$, which is in turn a function of time.
To illustrate how well the bias $V(v_m)$ compensates for the underlying FES $G(n)$, we can plot $(n(t),-V(v_m(t)))$ points over $G(n)$.
An example of such analysis is given in Figure~\ref{fig:panel-bias}.
The effective bias does not perfectly match $G(n)$, which is especially visible in the transition state region.
$v_m$ cannot sharply discriminate configurations with critical bubbles quite as well as $n$.
This is because $v_m$ is a \emph{global} order parameter, whereas bubble formation entails a \emph{local} density change.
As a result, configurations not directly around the top of $G(n)$ receive a bias meant for transition state configurations, and are therefore subjected to only a small biasing force.
This diminishes the ability of the bias to facilitate transitions.
However, the limited difference in barrier height on $G(n)$ and $G(v_m)$, respectively, means that the bias $V(v_m)$ does adequately compensate the FES on average.

\subsection{Superheated isotherm}

\begin{figure}[tb]
\includegraphics{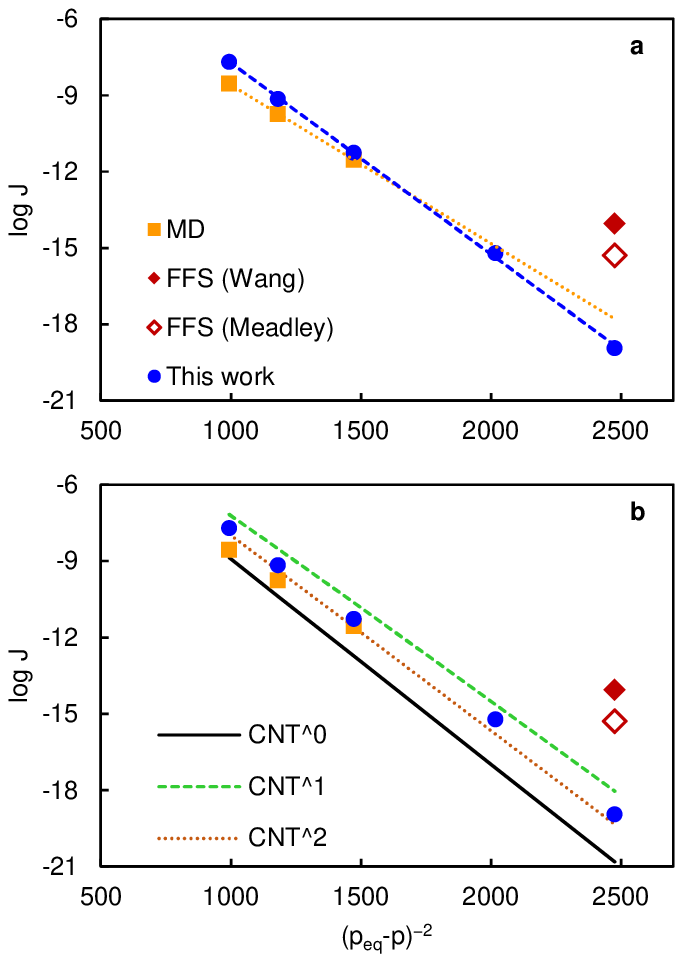}
\caption{\label{fig:panel-t855} Rate predictions from our approach and literature simulations, for the $T = 0.855$ superheated isotherm of the TFS--LJ fluid.
(a) Extrapolation of our data and MD rates using CNT relations.
(b) Comparing explicit rate predictions from simulation to CNT predictions based on different literature corrections to the surface tension (see text for details).
A first order correction employing a Tolman length $\delta_T = 0.128$ yields rates that overlap with the second order line.}
\end{figure}

\begin{table*}
\caption{\label{tab:t855} TST nucleation barriers $\Delta^\ddagger G$, extensive per-cell TST rates $k^\mathrm{TST}$, transmission coefficients $\kappa$ and final rate estimates $J$ for bubble nucleation on the $T = 0.855$ isotherm of the TFS--LJ fluid at different pressures $p$.\footnote{Reduced LJ units are used throughout.}
For comparison, predicted critical bubble radii $R_s^*$ and CNT barriers $G^*$ are reported for the ``best fit" CNT estimate, i.e., based on a second order correction to the surface tension.}
\begin{ruledtabular}
\begin{tabular}{ccccccc}
$p$     & $\Delta^\ddagger G$  & $k^\mathrm{TST}$           & $\kappa$        & $J$                            & $R_s^*$         & $G^*$ \\
\hline
0.01438 & $2.11 \pm 0.13$  & $1.15 \times 10^{-2 \pm 0.06}$ & $0.05 \pm 0.02$ & $1.99 \times 10^{-8 \pm 0.2}$  & $6.23 \pm 0.17$ & $13.96 \pm 0.65$ \\
0.01701 & $4.64 \pm 0.19$  & $5.97 \times 10^{-4 \pm 0.1}$  & $0.04 \pm 0.01$ & $7.11 \times 10^{-10 \pm 0.2}$ & $6.82 \pm 0.18$ & $16.77 \pm 0.78$ \\
0.02004 & $8.58 \pm 0.23$  & $5.99 \times 10^{-6 \pm 0.1}$  & $0.03 \pm 0.01$ & $5.45 \times 10^{-12 \pm 0.2}$ & $7.64 \pm 0.20$ & $21.15 \pm 0.98$ \\
0.02383 & $16.03 \pm 0.34$ & $9.80 \times 10^{-10 \pm 0.2}$	& $0.02 \pm 0.01$ & $6.21 \times 10^{-16 \pm 0.2}$ & $8.98 \pm 0.24$ & $29.40 \pm 1.37$ \\
0.02600 & $23.31 \pm 0.36$ & $1.96 \times 10^{-13 \pm 0.2}$ & $0.02 \pm 0.01$ & $1.14 \times 10^{-19 \pm 0.2}$ & $9.98 \pm 0.27$ & $36.40 \pm 1.69$ \\
\end{tabular}
\end{ruledtabular}
\end{table*}

We have probed five conditions in the superheated regime between $p = 0.01438$ and $p = 0.026$ on the $T = 0.855$ isotherm, well below the coexistence pressure $p_\mathrm{eq} = 0.0461$.
Nonequilibrium SMD simulation were carried out from $v_m = 1.7$ to $v_m = 2.2$ to parametrize $V (v_m)$ in cubic simulation cells containing $N = 17576$ atoms.
The computed barriers and rates are summarized in Table~\ref{tab:t855}.

Our rate estimates can be compared with previously published data for the same system.
On one end, Diemand et al.~\cite{Diemand2014} carried out MD simulations at low pressure ($p \leq 0.02383$) in very large simulation cells ($N \approx 5 \times 10^8$) in which several nucleation events could be observed directly.
On the other end, Wang et al.~\cite{Wang2009} and Meadley~\& Escobedo~\cite{Meadley2012} used forward flux sampling (FFS) to obtain rates at $p = 0.026$ in small systems of $N = 3375$ and $N = 8000$ atoms, respectively.
We have simulated the same pressures $p$ as these literature studies to allow for a direct one-to-one comparison.
The advantage of our approach is that it allows us the evaluate the nucleation rate over the full considered pressure range, so that disparate methodologies can now be compared to one consistent data set.

It can be seen from Figure~\ref{fig:panel-t855}a that our rate estimates closely match large-scale MD data at low $p$, but are orders of magnitude lower than FFS predictions at $p = 0.026$.
The FFS estimates of Wang et al. overshoot our values the most: As pointed out by Meadley~\& Escobedo, these rates are likely affected by finite size effects due to the small cell size that was employed in the FFS simulations.
Nevertheless, even Meadley~\& Escobedo's simulations in a larger cell produce a rate that is over two orders of magnitude higher than our estimate.

Rates from low-pressure MD simulations can, in principle, be extrapolated to higher pressures through CNT-derived relations.
According to CNT, $\ln J \sim (\Delta p)^{-2}$, in which $\Delta p$ is the pressure difference between the bubble and the surrounding liquid.
Assuming that $\Delta p$ is proportional to $p_\mathrm{eq}-p$, it thus becomes possible to extrapolate the MD results, as shown in Figure~\ref{fig:panel-t855}a.
The extrapolated MD results match our TST-based estimates within an order of magnitude over the whole considered range.
In addition, the TST-derived rates are very well-represented by a linear fit of $\log J$ to $(p_\mathrm{eq}-p)^{-2}$.

The applicability of CNT for bubble nucleation in superheated liquids has been controversial.~\cite{Wang2009,Diemand2014,Tanaka2015,Schmelzer2016}
Wang et al. analyzed temperature profiles in their FFS simulations.
They concluded that bubble formation is driven by the occurrence of local hot spots in the liquid.~\cite{Wang2009}
Because CNT (and also TST) are based on the assumption of local thermal equilibrium, such a fact would undermine the validity of the theory.
Diemand et al., however, observed that local hot spots never \emph{precede} bubble formation.
Rather, hot spots are a consequence of the excess kinetic energy that is produced by rapid expansion of the bubble.~\cite{Diemand2014}
Hot spots therefore only occur \emph{after} a larger-than-critical bubble is already formed.
The good cross-agreement between MD data, TST, and CNT corroborates this observation.

CNT in principle offers a path to nucleation rates that does not require any explicit simulation of the nucleation process.
Only a few bulk properties at coexistence are needed: the vapor/liquid surface tension $\gamma$, pressure $p_\mathrm{eq}$, liquid density $\rho_l$ and vapor density $\rho_g$.
Then, the CNT nucleation barrier $G^*$ is
\begin{equation}
  G^* = \frac{16 \pi}{3} \frac{\gamma^3}{(\Delta p)^2} ,
\end{equation}
in which $\Delta p$ is the pressure difference between the pressure inside and outside the bubble, which can be approximated though the Poynting correction as $\Delta p = (p_\mathrm{eq} - p) \delta_P$ in which
\begin{equation}
  \delta_P \approx 1 - \left( \frac{\rho_g}{\rho_l} \right) + \frac{1}{2} \left( \frac{\rho_g}{\rho_l} \right)^2 .
\end{equation}
At $T = 0.855$, $\rho_l = 0.729$ and $\rho_g = 0.0198$, so $\delta_P = 0.870$.
The nucleation rate can then be expressed as
\begin{equation}
  J = \rho_l \left( \frac{2 \gamma}{\pi m} \right)^{1/2} e^{-G^*/k_B T} . \label{eq:CNT}
\end{equation}

Both our TST-based approach and direct application of CNT rely on the concept of a free energy barrier on which the rates depends exponentially, supplemented by some prefactor.
In addition, both approaches assume that the system remains in local thermal equilibrium throughout the nucleation process, i.e., that there exists a time scale separation between bubble growth (slow) and atomic motion along other degrees of freedom (fast).

There are also two key differences.
First of all, the relation Eq.~\eqref{eq:CNT} can be derived from the explicit mechanistic assumption of a single bubble that grows or shrinks through evaporation or condensation of single atoms.
The TST equation Eq.~\eqref{eq:Eyring}, in contrast, merely describes the total flux through the dividing surface parametrized by the CV of choice; recrossings are explicitly accounted for by the committor analysis.
Second, the CNT barrier $G^*$ and TST barrier $\Delta^\ddagger G$ have different meanings.
The former is an intensive property, while the latter is extensive, as discussed before.~\cite{Yi2012,Bal2021JCP}
The extensive nature of $\Delta^\ddagger G$ therefore explains its lower absolute value in our large simulation cells, as can be seen in Table~\ref{tab:t855}.
$\Delta^\ddagger G$ is also reconstructed directly from an explicitly sampled thermodynamic ensemble of the simulated system, rather than computed from equilibrium properties only.
A direct comparison between the two approaches is therefore only possible on the basis of rates $J$.

Note that uncertainties in $\gamma$ (of about 20\%) have a rather large impact on $G^*$ and, therefore, result in CNT estimates of $J$ with an uncertainty of around one order of magnitude---larger than the uncertainties on our TST-based estimates (Table~\ref{tab:t855}).~\cite{Diemand2014}

A direct, zeroth order, application of CNT based on macroscopic input data yields fair agreement with our data and MD results, in line with observations of Diemand et al. (Figure~\ref{fig:panel-t855}b).
However, discrepancies remain.
The CNT rates can be further improved by taking into account the curvature dependence of surface tension $\gamma$.
A first order correction is given by the Tolman equation~\cite{Tolman1949}
\begin{equation}
  \gamma(R_s) = \frac{\gamma_\infty}{1 + 2 \delta_T / R_s} .
\end{equation}
Here, $R_s$ is the radius of the critical bubble or droplet, $\gamma_\infty$ the surface tension of a planar interface ($R_s \rightarrow \infty$), and $\delta_T$ the Tolman length.
$\delta_T$ is defined as the difference between the Gibbs equimolar radius $R_e$ and radius of the surface of tension $R_s$, although it is in practice an empirical parameter in the context of rate calculation.
Diemand et al. found a value of $\delta_T = 0.25$ to fit their MD data well.~\cite{Diemand2014}
Schmelzer~\& Baidakov later questioned the validity of a first order correction, especially for small $R_s$.~\cite{Schmelzer2016}
They argued that only a second order correction is able to accurately describe a wider range of curvatures:
\begin{equation}
  \gamma(R_s) = \frac{\gamma_\infty}{1 + 2 \delta_T / R_s + l^2 /R_s^2} ,
\end{equation}
where they proposed $\delta_T = 0.128$ and $l^2 = 1.56$.
To evaluate these corrections in the context of the rate we must know $R_s^*$, the radius of the critical nucleus, which CNT gives as $R_s^* = 2 \gamma (R_s^*) / \Delta p$.
We therefore obtain $\gamma (R_s^*)$ iteratively.

Application of the different corrections to $\gamma$ reveals their impact on the rate estimate, as can be seen in Figure~\ref{fig:panel-t855}b.
Setting $\gamma = \gamma_\infty = 0.0895$ results in an underestimation of the MD data and our results.
First order corrections bring the CNT estimate closer in absolute terms, although they appear to overcorrect somewhat.
Second order corrections then shift the rate back down.
This final data set has the closest agreement to the other rates estimates (except FFS) over the full pressure range.
Yet, the simple first order Tolman correction still appears to be adequate in this particular regime: If we set $\delta_T = 0.128$ (i.e., using the first order coefficient of the second order expansion) rate predictions are almost indistinguishable from the second order correction.
It appears that, at least under the present conditions, the good performance of the proposed second order correction can be attributed to its first order component.
For similar bubble sizes, Sanchez-Burgos et al. found a first order $\delta_T \approx 0.15$ from the extrapolation $\delta_T = \displaystyle \lim_{R_s \rightarrow \infty} (R_e - R_s)$.~\cite{SanchezBurgos2020}

\subsection{Cavitation isotherm}

\begin{figure}[tb]
\includegraphics{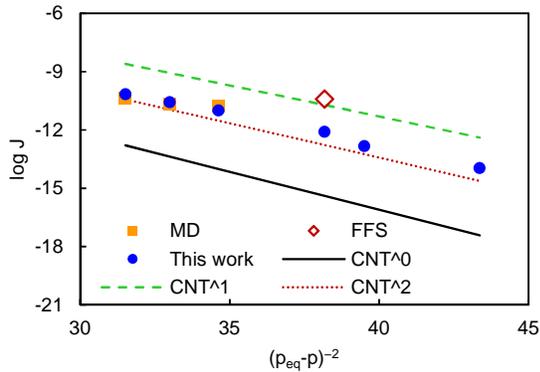}
\caption{\label{fig:panel-t700} Rate predictions from our approach and literature simulations, for the $T = 0.700$ cavitation isotherm of the TFS--LJ fluid.
Comparing explicit rate predictions from simulation to CNT predictions based on different literature corrections to the surface tension (see text for details).
A first order correction employing a Tolman length $\delta_T = 0.128$ yields rates that overlap with the second order line.}
\end{figure}

\begin{table*}
\caption{\label{tab:t700} TST nucleation barriers $\Delta^\ddagger G$, extensive per-cell TST rates $k^\mathrm{TST}$, transmission coefficients $\kappa$ and final rate estimates $J$ for bubble nucleation on the $T = 0.700$ isotherm of the TFS--LJ fluid at different pressures $p$.
Predicted critical bubble radii $R_s^*$ and CNT barriers $G^*$ are calculated as in Table~\ref{tab:t855}.}
\begin{ruledtabular}
\begin{tabular}{ccccccc}
$p$        & $\Delta^\ddagger G$  & $k^\mathrm{TST}$           & $\kappa$        & $J$                            & $R_s^*$         & $G^*$ \\
\hline
$-$0.16627 & $6.32 \pm 0.13$  & $1.35 \times 10^{-5 \pm 0.1}$  & $0.05 \pm 0.02$ & $6.66 \times 10^{-11 \pm 0.2}$ & $3.53 \pm 0.04$ & $16.02 \pm 0.34$ \\
$-$0.16222 & $7.01 \pm 0.23$  & $4.97 \times 10^{-6 \pm 0.1}$  & $0.06 \pm 0.03$ & $2.60 \times 10^{-11 \pm 0.3}$ & $3.62 \pm 0.04$ & $16.86 \pm 0.36$ \\
$-$0.15808 & $7.78 \pm 0.24$  & $1.67 \times 10^{-6 \pm 0.2}$  & $0.06 \pm 0.02$ & $9.74 \times 10^{-12 \pm 0.2}$ & $3.72 \pm 0.05$ & $17.79 \pm 0.37$ \\
$-$0.15000 & $9.66 \pm 0.23$  & $1.13 \times 10^{-7 \pm 0.1}$  & $0.08 \pm 0.03$ & $8.05 \times 10^{-13 \pm 0.2}$ & $3.92 \pm 0.05$ & $19.81 \pm 0.42$ \\
$-$0.14724 & $10.94 \pm 0.15$ & $1.82 \times 10^{-8 \pm 0.1}$  & $0.09 \pm 0.04$ & $1.46 \times 10^{-13 \pm 0.2}$ & $3.99 \pm 0.05$ & $20.57 \pm 0.43$ \\
$-$0.14000 & $13.07 \pm 0.14$ & $8.66 \times 10^{-10 \pm 0.1}$ & $0.14 \pm 0.05$ & $1.07 \times 10^{-14 \pm 0.2}$ & $4.19 \pm 0.05$ & $22.79 \pm 0.48$ \\
\end{tabular}
\end{ruledtabular}
\end{table*}

We considered six conditions on the $T = 0.700$ isotherm, where negative pressures are required to induce nucleation via cavitation and the coexistence pressure is $p_\mathrm{eq} =  0.01186$.
Nonequilibrium SMD simulation were carried out at pressures between $p = -0.16627$ and $-0.14000$ from $v_m = 1.4$ to $v_m = 1.6$.
Due to the smaller bubble sizes, cubic simulation cells with $N = 4096$ atoms were found to be sufficient.
The computed barriers and rates are summarized in Table~\ref{tab:t700}.

MD data are available for the lowest pressures ($p \leq -0.15808)$,~\cite{Diemand2014} and match our predictions well.
At higher pressures, only one FFS estimate is available, at $p = -0.15$.~\cite{Meadley2012}
The FFS rate overestimates our prediction by an order of magnitude, similar to the superheated case.
Our rate predictions appear to follow a CNT-style dependence on the pressure (Figure~\ref{fig:panel-t700}).

We verify the direct applicability of CNT-based formulas for rate calculation in this system.
Relevant parameters for this computation are $\rho_g = 0.0198$, $\rho_l =  0.729$ and $\gamma_\infty = 0.329$.
Assuming a curvature-independent $\gamma$ leads to a very large underestimation of the rate, an issue exacerbated in this system by the smaller bubble size (and larger curvatures) compared to the superheated system. 
A first order correction using $\delta_T = 0.25$ overcorrects, however.
It therefore appears that the second order correction is necessary, since it does lead to an excellent agreement with our data and MD results.
Yet, as in the superheated example, a first order correction with a smaller $\delta_T = 0.128$ performs almost identically.

\subsection{Superheated isobar}

\begin{figure}[tb]
\includegraphics{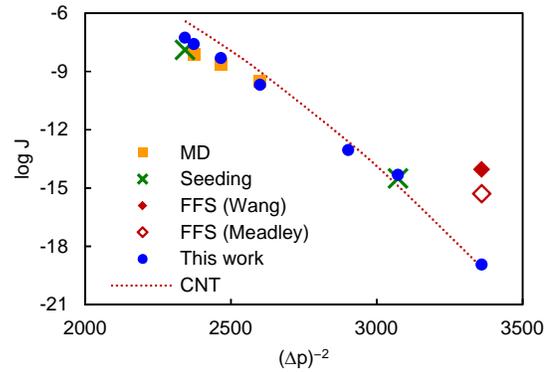}
\caption{\label{fig:panel-p026} Rate predictions from our approach and literature simulations, for the $p = 0.026$ superheated isobar of the TFS--LJ fluid.
Comparing explicit rate predictions from simulation to CNT predictions based on literature second order corrections to the surface tension.}
\end{figure}

\begin{table*}
\caption{\label{tab:p026} TST nucleation barriers $\Delta^\ddagger G$, extensive per-cell TST rates $k^\mathrm{TST}$, transmission coefficients $\kappa$ and final rate estimates $J$ for bubble nucleation on the $p = 0.026$ isobar of the TFS--LJ fluid at different temperatures $T$.
Predicted critical bubble radii $R_s^*$ and CNT barriers $G^*$ are calculated as in Table~\ref{tab:t855}, using empirical linear functions $\gamma_\infty(T)$ and $\Delta p (T)$.\footnote{The different calculation of $\gamma_\infty$ and $\Delta p$ explains the different numerical values and error bars for $T=0.855$ compared to Table~\ref{tab:t855}.}}
\begin{ruledtabular}
\begin{tabular}{ccccccc}
$T$        & $\Delta^\ddagger G$  & $k^\mathrm{TST}$               & $\kappa$        & $J$                            & $R_s^*$         & $G^*$ \\
\hline
0.8550     & $23.31 \pm 0.36$     & $1.96 \times 10^{-13 \pm 0.2}$ & $0.02 \pm 0.01$ & $1.14 \times 10^{-19 \pm 0.2}$ & $9.98 \pm 0.61$ & $35.89 \pm 3.79$ \\
0.8580     & $15.21 \pm 0.24$     & $2.74 \times 10^{-9 \pm 0.1}$  & $0.05 \pm 0.02$ & $4.64 \times 10^{-15 \pm 0.2}$ & $9.03 \pm 0.58$ & $27.72 \pm 3.10$ \\
0.8600     & $11.79 \pm 0.23$     & $1.53 \times 10^{-7 \pm 0.1}$  & $0.02 \pm 0.01$ & $9.00 \times 10^{-14 \pm 0.2}$ & $8.43 \pm 0.57$ & $23.24 \pm 2.71$ \\
0.8640     & $5.65 \pm 0.16$      & $1.99 \times 10^{-4 \pm 0.1}$  & $0.03 \pm 0.01$ & $2.03 \times 10^{-10 \pm 0.2}$ & $7.34 \pm 0.54$ & $16.14 \pm 2.05$ \\
0.8660     & $3.40 \pm 0.10$      & $2.73 \times 10^{-3 \pm 0.1}$  & $0.05 \pm 0.02$ & $4.86 \times 10^{-9 \pm 0.2}$  & $6.83 \pm 0.52$ & $13.35 \pm 1.77$ \\
0.8675     & $1.80 \pm 0.14$      & $1.73 \times 10^{-2 \pm 0.1}$  & $0.04 \pm 0.02$ & $2.51 \times 10^{-8 \pm 0.2}$  & $6.46 \pm 0.51$ & $11.54 \pm 1.59$ \\
0.8680     & $1.44 \pm 0.20$      & $2.64 \times 10^{-2 \pm 0.1}$  & $0.06 \pm 0.02$ & $5.39 \times 10^{-8 \pm 0.2}$  & $6.35 \pm 0.51$ & $10.98 \pm 1.53$ \\
\end{tabular}
\end{ruledtabular}
\end{table*}

Rosales-Pelaez et al. studied nucleation on the $p=0.026$ isobar using direct brute force MD simulations as well as a seeding approach.~\cite{RosalesPelaez2019}
We consider temperatures within the same range, in casu between $T = 0.855$ and $T = 0.868$.
These conditions intersect with our work on the superheated isotherm for $(p,T) = (0.026,0.855)$, for which FFS results are also available.~\cite{Wang2009,Meadley2012}
Likewise, initial nonequilibrium sampling was carried out from $v_m = 1.7$ to $v_m = 2.2$ in cubic simulation cells containing $N = 17576$ atoms.
The computed barriers and rates are summarized in Table~\ref{tab:p026}.

Rosales-Pelaez et al. produced two sets of rate estimates from seeding, which differed in how the critical bubble radius was defined.
They noted that neither set could simultaneously match their MD results at high temperatures, and literature FFS data at $T = 0.855$.
As it turns out, seeding predictions based on the Gibbs dividing surface (GDS)---which Rosales-Pelaez et al. found to be consistent with MD simulations at low temperature---are further validated by our data (Figure~\ref{fig:panel-p026}).
While the FFS estimates at $T = 0.855$ overshoot our rate prediction by orders of magnitude, as noted earlier, the GDS-based rate derived from seeding at $T = 0.858$ agrees very closely with our prediction.

From a CNT point of view, the pressure-dependence of rate is primarily impacted by the change in $\Delta p$.
The effect of a change in temperature is somewhat more complex to account for, since $\gamma$ and $\rho_g$ and $\rho_l$ are also temperature-dependent.
This means that extrapolation of nucleation rates to lower $T$ is more difficult.
We can use the empirical relations $\gamma_\infty (T) = 1.3557 - 1.4809 T$ and $\Delta p (T) = -0.2071 + 0.2624 T$.~\cite{RosalesPelaez2019} 
Compared to literature, we have shifted the $\gamma_\infty (T)$ curve to match the value of $\gamma_\infty$ at $T = 0.855$ reported by Diemand et al.~\cite{Diemand2014} and allow for consistency with our simulations on the $T = 0.855$ isotherm.
If we further assume $\rho_g$ and $\rho_l$ to be constant in the considered temperature range, and use the second-order correction to $\gamma$, we can plot CNT-predicted rates in Figure~\ref{fig:panel-p026}.
Although a linear fit of our $\log J$ data to $(\Delta p)^2$ might seem appropriate within the studied range, the nonlinear behavior of the CNT rate reveals that linear extrapolation far beyond the simulated temperature range is not advisable.
The uncertainties on the $\gamma_\infty(T)$ fit are also fairly high, which affects the fidelity of CNT for rate predictions (compare $R_s^*$ and $G^*$ in Tables~\ref{tab:t855} and \ref{tab:p026}).
If accurate temperature-dependent surface tensions and coexistence pressures are known, however, a straightforward application of CNT accurately describes the nucleation rate.

\subsection{General remarks}

Our results allow us to reconcile rate predictions from direct MD simulations, seeding approaches, and CNT.
By combining a modern, efficient free energy method with a rate evaluation based on a generic implementation of TST, we are able to span the full range of rates and conditions that were previously reported within a single, consistent approach that is mostly free of mechanistic assumptions.
It therefore becomes easier to cross-validate other rate computation approaches, and to investigate the applicability of extrapolation techniques.
We note that our approach was already validated for droplet nucleation processes, highlighting its general applicability.~\cite{Bal2021JCP}

One consistent outlier has been forward flux sampling (FFS), yielding bubble nucleation rates that are consistently above the range established by other methods by several orders of magnitude.
Finite size effects can explain some of the inconsistencies,~\cite{Meadley2012} but not all; the FFS rate overestimation persists even for the cavitation process when using larger cells than in our simulations.~\cite{Meadley2012}
Large discrepancies, up to seven orders of magnitude, between FFS and other methods have also been reported for ice nucleation.~\cite{Cheng2018}
Given that FFS has become a workhorse tool for the computation of nucleation rates in diverse systems,~\cite{Hussain2020} it might be prudent to further validate the method and investigate the origins of such large deviations.
That being said, certain deficiencies of FFS in the context of nucleation have already been addressed in later developments of the method.~\cite{HajiAkbari2018}

Cross-method benchmarking, reproducibility, and accuracy of nucleation rate prediction has seen a renewed interest, as recently discussed in an excellent review.~\cite{Blow2021} 
We hope our benchmarking strategies and methodological insights can also contribute to this discussion.

We note that the methodology as outlined in in Sec.~\ref{sec:method} is not completely set in stone.
Its only fixed aspects are the reconstruction of the nucleation free energy surface $G(\chi)$ along a suitable order parameter $\chi$---which can be fed into the TST expressions Eqs.~\eqref{eq:Eyring}--\eqref{eq:Gl}---followed by committor analysis and the recrossing correction of Eq.~\eqref{eq:committor}.
There exists a wide variety of methods that can reconstruct free energies, each with their own characteristics and strengths.~\cite{Henin2022}
The particular choice of free energy method in practice mostly depends on the problem at hand, the user's expertise, and code availability.
We eat our own cooking by using reweighted Jarzynski sampling, but are also motivated by its good sampling efficiency and convergence in head-to-head comparisons with established adaptive bias methods.~\cite{Bal2021JCTC}
The specific choice of reaction coordinate $\chi$ will be system-dependent and can take much more sophisticated forms in studies of crystallization.~\cite{Piaggi2020,Zou2021,Karmakar2021}
As noted before, sampling efficiency and accuracy of rate is highly dependent on an appropriate choice of this reaction coordinate.
That being said, the choice of $\chi$ is the \emph{only} system-specific aspect of the overall procedure.
Our rate calculation procedure is therefore as generic and process-agnostic as the free energy methods on which it is based, equally applicable to nucleation as it is to chemical reactions.~\cite{Bal2021JCP}

Finally, we have confirmed the validity of CNT for the considered conditions.
First, relations derived from CNT allow for the extrapolation of explicit rate predictions to conditions that are more difficult to sample.
Second, if high-quality equilibrium properties of the liquid are known, it is possible to produce accurate rate estimates with CNT.
An appropriate curvature correction to the surface tension is however required.
Although we find that a first order Tolman correction is sufficient in our case, it is possible that second order corrections are needed for a consistent description of even larger curvatures, i.e., in the case of higher superheatings (or supercoolings in case of droplet nucleation).~\cite{Schmelzer2016}
A recent study demonstrated that a consistent simultaneous treatment of bubble and droplet nucleation is possible using the same first order Tolman correction, although only systems with fairly large bubble/droplet sizes were fitted.~\cite{SanchezBurgos2020}

\section{Conclusions}

We have revisited a paradigmatic nucleation problem---rate calculation of bubble formation in a Lennard-Jones fluid---with a set of recently developed enhanced sampling methodologies based on the Jarzynski equality, transition state theory, and a simple recrossing correction.
The employed methodology is highly generic---the only system- or process-dependent aspect being the choice of an approximate reaction coordinate.
A very wide range of simulation conditions can thus be consistently probed within an identical simulation setup.
We were therefore able to reconcile disparate literature data on the same system.

On one hand, we validate simulated nucleation rate predictions based on brute force molecular dynamics and seeding approaches, while forward flux sampling (FFS) appears to produce outliers.
We also show that simple analytical expressions derived from classical nucleation theory can produce satisfactory rate predictions, provided that the curvature dependence of the surface tension is empirically corrected.
Another viable approach is to extend explicit rate predictions to a wider range of conditions with extrapolation schemes that exploit CNT trends.

On the other hand, our results confirm that our generic TST-based approach to rate calculation is equally applicable to complex nucleation processes as it is to simple chemical reactions.
It may therefore be a useful additional tool for rate calculation in future modeling studies, either to be used on its own or for cross-validation purposes.
We therefore hope that our insights can help improve the accuracy and reproducibility of rate calculation also for other types of (nucleation) processes.

\begin{acknowledgments}
K.M.B. was funded as a junior postdoctoral fellow of the FWO (Research Foundation -- Flanders), Grant 12ZI420N.
The computational resources and services used in this work were provided by the HPC core facility CalcUA of the Universiteit Antwerpen, and VSC (Flemish Supercomputer Center), funded by the FWO and the Flemish Government.
\end{acknowledgments}

\section*{Author Declarations}
\subsection*{Conflict of interest}
The authors have no conflicts to disclose.

\section*{Data availability}
The data that support the findings of this study are available from the corresponding author upon reasonable request.
Sample inputs to reproduce the reported simulations are deposited on PLUMED-NEST (www.plumed-nest.org), the public repository of the PLUMED consortium~\cite{PLUMED2019}, as plumID:22.025.\cite{data}

\bibliography{bibliography}

\end{document}